\documentstyle[graphicx,float]{article}

\begin{document}
\title{Entanglement in Absorption Processes}
\date{}
\author{Pedro Sancho \\ Centro de L\'aseres Pulsados CLPU \\ Parque
Cient\'{\i}fico, 37085 Villamayor, Salamanca, Spain}
\maketitle
\begin{abstract}
Entanglement can modify light-matter interaction effects and,
conversely, these interactions can change the non-classical
correlations present in the system. We present an example where
these mutual connections can be discussed in a simple way at the
graduate and advanced undergraduate levels. We consider the process
of light absorption by multi-atom systems in non-product states,
showing first that the absorption rates depend on entanglement. The
reverse is also true, absorption processes can generate an
hyperentangled atomic state involving in a non-product form both
internal and spatial variables. This behavior differs from that of
spontaneous emission, which disentangles atomic systems.
\end{abstract}


\section{Introduction}

Light-matter interaction effects can be modified by the presence of
entanglement. These modifications have been studied in several
contexts \cite{fic,dow,fri,fed}. In particular, absorption and
emission rates have been analyzed in detail \cite{yo}. In the case
of spontaneous emission two experiments \cite{jap,bel} have tested
these effects (see \cite{yop} for a discussion of the experiments
interpretation).

Some examples of the reverse process, variations of the entanglement
present in multi-atom systems due to light-matter interactions, have
also been described in the literature. For instance, spontaneous
emission disentangles initially correlated atomic systems
\cite{ebe,eb1}. Another two well-know examples are the interaction
of trapped ions with laser beams \cite{cir} and of atoms with the
field in a cavity \cite{bge}, where the light absorption generates
entanglement in the ionic and atomic systems.

We consider an example, light absorption by two-atom systems, that
presents the above ideas in a simple way. The calculations and
concepts involved are not difficult and could be used at a graduate
and advanced undergraduate level to introduce the subject. Moreover,
our presentation highlights a fundamental difference between
absorption and spontaneous emission: the first can change atomic
non-classical correlations whereas the second destroys them.

In the example analyzed here the atomic state is initially only
entangled in the spatial variables, but after the absorption it
becomes hyperentangled, involving the correlations both spatial and
internal degrees of freedom. By evaluating the von Neumann entropy
we show that the degree of entanglement does not change, but it is
redistributed between the two degrees of freedom. The absorption
process can generate hyperentanglement without changing the initial
degree of entanglement. The evaluation of the von Neumann entropy
involves non-orthogonal states. As these calculations are simple
this example could also be useful to teach at a graduate level how to evaluate
entanglement when the states overlap, a situation frequently found in atomic and
molecular physics.

\section{Modification of absorption rates}

We discuss in this section how entanglement changes the absorption
rates in multi-atom systems. First, we describe the arrangement. A
source prepares pairs of distinguishable atoms in a non-factorizable
state traveling in opposite directions. The preparation of entangled
states is not a simple task. In the case of photons there are
several well-developed techniques as spontaneous parametric
down-conversion, or those based on quantum dots in semiconductors
and nanoscale impurities in diamonds. In the atomic case there is
not a so vast literature on the generation issue. For our proposal
we can invoke the experiments \cite{jap,bel}, based on molecular
photodissociation (although at variance with the two above
references, the decaying atoms must be in their ground states
instead of excited ones). Because of momentum conservation the two
atoms travel in almost opposite directions. The state representing
this preparation is
\begin{eqnarray}
|\psi _0>=\frac{1}{\sqrt 2}(|\phi _L>_A|g>_A |\varphi _R>_B|g>_B + |\varphi
_L>_B|g>_B |\phi _R>_A|g>_A) = \nonumber \\
\frac{1}{\sqrt 2}(|\phi _L>_A|\varphi _R>_B + |\varphi _L>_B |\phi
_R>_A)|g>_A|g>_B
\end{eqnarray}
The labels $A$ and $B$ refer to the two atoms, and $L$ and $R$ denote
opposite traveling directions. The symbol $\phi _L $ ($\varphi _R $)
represents the center of mass (CM) wave function of atom $A$ ($B$)
moving towards $L$ ($R$). On the other hand, $|g>_i,i=A,B$
represents the electronic ground state of the atom $i$.

When the separation of the atoms is large, with no spatial overlap
between them, they interact with the light. We consider classical
light in the linear (without multiple absorptions) regime. In order
to the atomic correlations be able to modify absorption rates it is
not necessary to consider more sophisticated types of light (quantum
or entangled) than the classical one, or to move to the non-linear
regime. The beams must contain the absorption frequencies of the two
atoms; we can use light beams with different frequencies or a single
broad band beam. As we assume a low intensity of the beams (linear
regime), after the interaction the atomic states evolve as
\begin{equation}
|\phi _j >_A|g>_A \rightarrow \alpha |\bar{\phi}_j>_A |e>_A + \beta |\phi _j>_A
|g>_A
\label{eq:dos}
\end{equation}
and
\begin{equation}
|\varphi _j>_B|g>_B \rightarrow \gamma |\bar{\varphi}_j>_B |e>_B + \delta
|\varphi _j>_B |g>_B
\label{eq:tre}
\end{equation}
with $j=L,R$. The coefficients obey the relations $|\alpha |^2
+|\beta |^2=1$ and $|\gamma |^2 + |\delta |^2=1$. The wave functions
$\bar{\phi}_j$ and $\bar{\varphi}_j$ include the recoil after the
absorption, and $|e>$ denotes the excited internal state.

\begin{figure}[H]
\center
\includegraphics[width=12cm,height=2.5cm]{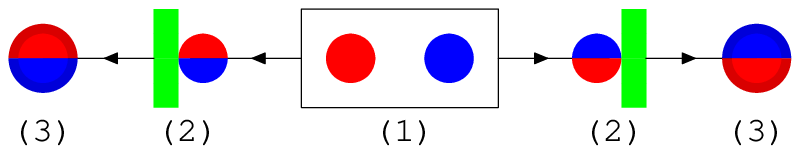}
\caption{Sketch of the arrangement. The temporal evolution consists of three
stages: (1) Preparation of an entangled state for atoms A and B, here
denoted by the red and blue colors. The box represents the device or
procedure used in the preparation. (2) The entangled state evolves
freely moving away from the source. We do not know which atom goes
to the left and which one to the right because we have a
superposition of the two-atom alternatives $|red>_L|blue>_R$ and
$|blue>_L|red>_R$. This superposition is graphically represented by
the circles doubly colored (the two half circles over the evolution
line represent the first alternative,...). At the end of this stage the
atoms interact with light beams drawn as green rectangles. (3) After
the light-matter interaction the atoms can become excited.}
\end{figure}

Note that the Eqs. (\ref{eq:dos}) and (\ref{eq:tre}) are only valid
in the low-intensity beam approximation. This approximation is
quantitatively expressed by the relations $|\alpha|^2 \ll 1$ and
$|\gamma|^2 \ll 1$, indicating that only single absorptions are
relevant in the problem. The probability of multi-absorption
processes is very low and can be neglected. In the presence of
multi-photon absorptions we enter in a non-linear regime and new
terms should be added to these equations.

The final state after the interaction is
\begin{equation}
|\psi _f>= \frac{1}{\sqrt 2} ( \alpha \gamma  |\bar{\phi}_L>_A
|\bar{\varphi}_R>_B + \gamma \alpha  |\bar{\phi}_R>_A
|\bar{\varphi}_L>_B) |e>_A|e>_B + |\cdots>
\end{equation}
where $|\cdots >$ contains the rest of terms, which do not lead to
double absorptions.

We represent the above arrangement in Fig. 1.

From the expression for $\psi _f$ we can calculate the probabilities
of double absorptions. There are two alternatives that contribute to
the probability of double absorption, (i) absorption by an atom of
type $A$ in $L$ and by one of type $B$ in $R$ and, (ii) absorption
by an atom of type $B$ in $L$ and by one of type $A$ in $R$. These
alternatives are not distinguishable (both atoms can absorb at both
sides of the arrangement) and consequently, according to the rules
of quantum theory, we must add probability amplitudes instead of
probabilities. Finally, the probability of double absorption is
\begin{equation}
P_{dou}=\left| \frac{1}{\sqrt 2} \alpha \gamma   + \frac{1}{\sqrt 2}
\gamma \alpha  \right|^2 = 2|\alpha \gamma |^2
\end{equation}

The above probability differs from that of atoms in product states.
In effect, when the initial atomic state is, instead of the pure one
$\psi _0$, a mixture of $|\phi _L>_A|g>_A |\varphi _R>_B|g>_B$ and
$|\phi _R>_A|g>_A |\varphi _L>_B|g>_B$ with equal weights $1/2$, the
double absorption probability changes to
\begin{equation}
P_{dou}^{mix}= \frac{1}{2} |\alpha \gamma |^2+ \frac{1}{2} |\gamma
\alpha|^2  = |\alpha \gamma |^2
\end{equation}
The probability of double absorption in the entangled state doubles
that in product ones. We conclude that the absorption probabilities in
multi-atom systems depend on entanglement.

The experimental verification of the above ideas relies on the
quantum theory of detection. In our case its implementation is
simple. We do not need to measure the atoms and their internal
states. As an excited atom shortly emits a photon because of
spontaneous emission we only need standard optical detectors. The
presence or absence of photon detections at every side of the
arrangement tell us if the atoms were excited or not.

\section{Modification of entanglement}

In this section we describe the reverse of the above behavior, showing that the
absorption process can also modify the entanglement distribution of
the atomic system. The explicit expression for $|\cdots >$ is
\begin{eqnarray}
\sqrt 2 |\cdots> = \alpha \delta (|\bar{\phi }_L>_A|\varphi _R>_B+ |\bar{\phi
}_R>_A |\varphi _L>_B)|e>_A |g>_B + \nonumber \\
\beta \gamma (|\phi _L>_A|\bar{\varphi }_R>_B+ |\phi _R>_A |\bar{\varphi
}_L>_B)|g>_A |e>_B + \nonumber \\ \beta \delta
(|\phi _L>_A |\varphi _R>_B + |\phi _R>_A |\varphi _L>_B)|g>_A |g>_B )
\end{eqnarray}
From this expression it is immediate to see that $\psi _f$ is
entangled, but in a very different way from $\psi _0$. The final
state is hyperentangled. Hyperentanglement refers to entanglement
involving more than one degree of freedom. In our case we clearly
have hyperentanglement as the CM and internal degrees are involved.
In general, the hyperentanglement studied in the literature has the
form of entangled states in each one of the degrees of freedom, and
all of these states in product form (in our arrangement would be
$(|\phi _L>_A |\varphi _R>_B + |\phi _R>_A |\varphi _L>_B)(|e>_A
|g>_B + |g>_A |e>_B)$). For $\psi _f$ is no longer possible to
express the state as a product of the CM and internal parts (as it
was the case for $\psi _0$).

Initially we only had correlations in the CM variables. After the
interaction we have entanglement between the two variables, a form
clearly different from the initial one. We conclude that the process
of absorption has modified the entanglement of the system,
generating hyperentanglement.

We can associate the generation of hyperentanglement with the recoil
of the atom. If we neglect the effect of the recoil we can make the
approximation $\bar{\phi} \approx \phi$ and $\bar{\varphi} \approx
\varphi$. Then the final state can be approximated as (using Eqs.
(\ref{eq:dos}) and (\ref{eq:tre}) without recoil)
\begin{eqnarray}
|\psi_f^{approx}>=\frac{1}{\sqrt 2}(|\phi _L>_A |\varphi _R>_B +
|\phi _R>_A |\varphi _L>_B) \times \nonumber \\
(\alpha |e>_A +\beta |g>_A) (\gamma |e>_B +\delta |g>_B)
\end{eqnarray}
where, as in the initial state, there is only entanglement between the two
particles in the spatial variables.

If the initial state in not entangled the situation will change
drastically. Imagine that initially we have the product state $|\phi
_L>_A|g>_A |\varphi _R>_B|g>_B$. It changes to $(\alpha |\bar{\phi}
_L>_A|e>_A + \beta |\phi _L>_A|g>_A)(\gamma |\bar{\varphi
}_R>_B|e>_b + \delta |\varphi _R>_B|g>_B)$ after the interaction
with the light. This is also in a product form and, consequently,
the absorption does not generate entanglement between $A$ and $B$.
Thus, the initial presence of some degree of entanglement in the
atomic system (in addition to the process of recoil) is a necessary
condition for the generation of hyperentanglement in our
arrangement.

In the above paragraphs we have only given a qualitative description
of the problem. In the next section we address the subject from a
more quantitative point of view. We evaluate the initial and final
degrees of freedom showing that there is not generation of
entanglement but only a redistribution between different degrees of
freedom.

\section{Evaluation of the entanglement degree}

We evaluate the entanglement degree after the absorption of the
photons. We use the von Neumann entropy as measure of the
entanglement degree. The evaluation is a little bit involved because
after the recoil the states $\bar{\phi}$ and $\bar{\varphi}$ overlap
with $\phi$ and $\varphi$. In order to correctly deal with that
overlap we introduce states orthogonal to the last ones, $\phi
^{\bot}$ and $\varphi ^{\bot}$, such that the first ones can be
expressed as
\begin{equation}
|\bar{\phi }_i>=a|\phi _i> + b|\phi _i^{\bot}>\end{equation}
and
\begin{equation}
|\bar{\varphi }_i>=c|\varphi _i> + d|\varphi _i^{\bot}>
\end{equation}
with $i=L,R$. We assume by the matter of simplicity $a,b,c,d$ to be
real, obeying the relations $a^2+b^2=1$ and $c^2+d^2=1$. In
\cite{sr} a similar calculation, in a different context, has been
carried out to determine the entanglement degree of overlapping
states.

To calculate the von Neumann entropy we need to determine the
eigenvalues of the reduced density matrix or, equivalently, the
coefficients of the Schmidt form of the state $\psi _f$ \cite{fri}. In the last
approach we express the state in the matrix form
\begin{equation}
\Lambda \equiv \left(
\begin{array}{cl}
0 & \tilde{\Lambda} \\
\tilde{\Lambda} & 0
\end{array}
\right)
\end{equation}
with
\begin{equation}
\tilde{\Lambda} \equiv \left(
\begin{array}{clcr}
 \beta \delta & \alpha \delta a & \alpha \delta b \\
\beta \gamma c & \alpha \gamma ac & \alpha \gamma bc \\
\beta \gamma d & \alpha \gamma ad & \alpha \gamma bd
\end{array}
\right)
\end{equation}
where the matrix $\Lambda$ is written in the basis $|\phi _L,g>_A,
|\phi _L,e>_A, |\phi _L^{\bot},e>_A, |\phi _R,g>_A, |\phi _R,e>_A,
|\phi _R^{\bot},e>_A $ for the particle $A$ (rows) and $|\varphi _L,g>_B,
|\varphi _L,e>_B, |\varphi _L^{\bot},e>_B, |\varphi _R,g>_B |\varphi
_R,e>_B, |\varphi _R^{\bot},e>_B $ for $B$ (columns). We have not included the
normalization factor $1/\sqrt 2$ in these expressions because it
will disappear in the final normalization of the diagonalized state.

Now, the Schmidt form can be obtained from the diagonalization of
$\Lambda$. The coefficients of the Schmidt form correspond to the
eigenvalues of the matrix, given by the solutions of
\begin{equation}
det(\Lambda - \lambda \hat{I}) =0
\end{equation}
where $\hat{I}$ is the $6 \times 6$ identity matrix. The explicit form of this
equation is
\begin{equation}
\lambda ^6 -\lambda ^4 (\alpha \gamma (ac+bd)+ \beta \delta) ^2 =0
\end{equation}
with quadruple null solution $\lambda ^4=0$ and
\begin{equation}
\lambda _{\pm}= \pm (\alpha \gamma (ac+bd)+ \beta \delta)
\end{equation}
Finally, the normalized diagonalized state reads
\begin{eqnarray}
|\psi _f>= \frac{\lambda _+}{\sqrt {\lambda _+^2 + \lambda _-^2}} |\chi
_+>|\bar{\chi}_+> + \frac{\lambda _-}{\sqrt {\lambda _+^2 + \lambda _-^2}} |\chi
_->|\bar{\chi}_->= \nonumber \\ \frac{1}{\sqrt 2} |\chi _+>|\bar{\chi}_+> -
\frac{1}{\sqrt 2} |\chi _->|\bar{\chi}_->
\label{eq:vne}
\end{eqnarray}
with $\chi$ and $\bar{\chi}$ denoting the eigenkets associated with these
eigenvalues.

Now, we are in position to evaluate the von Neumann entropy. We have
$S=- (1/2) \log _2 1/2 - (1/2) \log _2 1/2 =1$. The entropy of the
final state is $1$, as it can be also directly derived from Eq.
(\ref{eq:vne}), corresponding to the maximally entangled form of a
qubit.

With respect to the initial atomic state, the form of the spatial
variables is also that of a maximally entangled qubit. On the other
hand, the state of the internal variables is a product one and does
not contribute to the entanglement degree. Consequently, the initial
entanglement is also $S=1$. The degree of entanglement does not
change because of the light absorption. The entanglement has
been only redistributed between more degrees of freedom.

\section{Discussion}

We have analyzed in a simple example the mutual dependence between
entanglement and absorption. This is a two-sided relation. The
modifications associated with the presence of entanglement on the
absorption/emission rates of multi-atom systems have been
extensively described in the literature. We have emphasized the much
more asymmetric behavior of the reverse effects. Light-matter
interactions can change (in the case of absorption) or destroy (for
spontaneous emission) non-classical correlations.

It is also instructive to present the process of entanglement
redistribution in terms of the LOCC (Local Operations and Classical
Communication) paradigm. This is a central element in the
characterization of entanglement measures \cite{inf}. According to
it, physical operations that only affect to one of the components of
a multi-particle system, and information transmitted between
different components via classical means cannot modify the
entanglement degree of the full system. This is our case. As the
photon absorptions occur at well-separated places they are local
operations and the process is within the LOCC paradigm. The total
degree of entanglement does not change. However, the paradigm says
nothing about how this entanglement is shared among the different
variables, allowing for the redistribution.

We have been only concerned with the entanglement behavior in the
atomic system. The light beams have been considered as classical
auxiliary tools and we have not take care about their properties (we
have only demanded them not to be very intense and to contain the
adequate frequencies). A potential extension of the work would be to
study similar processes with quantum and entangled light. As it is
well-known the extension can offer advantages over the classical
framework. For instance, in \cite{Sal} the authors described an
increase of exciton oscillator strengths in absorption processes by
semiconductor quantum wells. Similarly, the Fourier limitations on
spectral resolution can be circumvented \cite{Dor} and the
antibunching effects improved \cite{JCL}.

Although the main aim of the paper is pedagogical we must briefly
consider potential  applications of the scheme. In general, the
entanglement generated this way is short lived because it quickly
disappears with the subsequent spontaneous emission. Nevertheless,
there is a scenario where it is possible, in principle, to exploit
it. This potentially interesting scenario considers transitions to
metastable excited states. In this case, we have the possibility of
manipulating the system during an interval of time long enough
(similar to those of some processes in ion trapping) to try of
exploiting the resource. For instance, we could study if the
non-product form of hyperentaglement described here differs from the
standard product one. It could be also on the basis of schemes to
distribute pre-existing entanglement between various degrees of
freedom initially not correlated.


\begin{thebibliography}{99}
\bibitem{fic} Ficek Z and Tana\'{s} R 2002 Entangled states and collective
nonclassical effects in two-atom systems {\it Phys. Rep.} {\bf 372} 369
\bibitem{dow} Dowling J P 2008 Quantum optical metrology-the lowdown on high-
N00N states {\it Contem. Phys.} {\bf 49} 125
\bibitem{fri} Tichy M C, Mintert F and Butchleitner A 2011 Essential
entanglement for atomic and molecular physics {\it J. Phys. B} {\bf 44} 192001
\bibitem{fed} Fedorov M V, Efremov M A, Kazakov A E, Chan K W, Law C K and
Eberly J H 2005 Spontaneous emission of a photon: Wave-packet structures and
atom-photon entanglement {\it Phys. Rev. A} {\bf 72} 032110
\bibitem{yo}  Sancho P 2016 Atomic absorption and emission in non-product states
{\it Eur. Phys. J. D} {\bf 70} 188
\bibitem{jap} Tanabe T, Odagiri T, Nakano M, Kumagai Y, Suzuki I H, Kitajima M
and Kouchi N 2010 Effect of entanglement on the decay dynamics of a pair of
$H(2p)$ atoms due to spontaneous emission {\it Phys. Rev. A} {\bf 82}  040101(R)
\bibitem{bel} Urbain X, Dochain A, Lauzin C and Fabret B 2015 Absence of
entanglement effect on the decay dynamics of $H(2p)$ pairs produced by VUV
photodissociation of $H_2$ {\it J. Phys. Conf. Ser.} {\bf 635} 112085
\bibitem{yop} Sancho P 2017 Entanglement, identity, and disentanglement in two-
atom spontaneous emission {\it Phys. Rev. A} {\bf 95} 032116
\bibitem{ebe} Yu T and Eberly J H 2004 Finite-time disentanglement via
spontaneous emission {\it Phys. Rev. Lett.} {\bf 93} 140404
\bibitem{eb1} Yu T and Eberly J H 2009 Sudden death of entanglement {\it
Science} {\bf 323} 598
\bibitem{cir} Zoller P, Cirac J I, Duan L and Garc\'{\i}a-Ripoll JJ 2004
Implementing quantum information processing with atoms, ions and photons {\it
Proceedings of the Les Houches Summer School 2003}
\bibitem{bge} Walther H, Varcoe B, Englert B-G and Becker T 2006 Cavity Quantum
Electrodynamics {\it Rep. Prog. Phys.} {\bf 69} 1325
\bibitem{sr} Lo Franco R and Compagno G 2016 Quantum entanglement of identical
particles by standard information-theoretic notions {\it Sci. Rep.} {\bf 6}
20603
\bibitem{inf} Plenio M B and Virmani S 2007 An introduction to entanglement
measures {\it Quant. Inf. Comput.} {\bf 7} 1
\bibitem{Sal} Salazar L J, Guzman D A, Rodr\'{\i}guez F J and Quiroga L 2012
Quantum-correlated two-photon transitions to excitons in semiconductor quantum
wells {\it Optics Express} {\bf 20} 4770
\bibitem{Dor} Dorfman K E, Schlawin F and Mukamel S 2016 Nonlinear optical
signals and spectroscopy with quantum light {\it Rev. Mod. Phys.} {\bf 88}
045008
\bibitem{JCL} L\'opez Carre\~{n}o J C, S\'anchez Mu\~{n}oz C, del Valle E and
Laussy F P 2016 Excitation with quantum light II. Exciting a two-level system
{\it Phys. Rev. A} {\bf 94} 063826
\end{thebibliography}
\end{document}